\documentstyle[amsfonts,aps,12pt]{revtex}

\begin{document}
\author{G. H. Rawitscher}
\title{Inclusion of virtual nuclear excitations in the formulation of the (e,e'N)
reaction.}
\address{Physics Department, University of Connecticut, Storrs, CT 06268\bigskip\\
bigskip\ }
\maketitle

\begin{abstract}
\ \ A wave-function framework for the theory of the (e,e'N) reaction is
presented in order to justify the use of coupled channel equations in the
usual Feynman matrix element. The overall wave function containing the
electron and nucleon coordinates is expanded in a basis set of eigenstates
of the nuclear Hamiltonian, which contain both bound states as well as
continuum states.. The latter have an ingoing nucleon with a variable
momentum $Q$ incident on the daughter nucleus as a target, with as many
outgoing channels as desirable. The Dirac Eqs. for the electron part of the
wave function acquire inhomogeneous terms, and require the use of distorted
electron Green's functions for their solutions. The condition that the
asymptotic wave function contain only the appropriate momentum $Q_k$ for the
outgoing nucleon, which corresponds to the electron momentum $k$ through
energy conservation, is achieved through the use of the steepest descent
saddle point method, commonly used in three-body calculations. \bigskip\
\newpage\ 
\end{abstract}

\section{\ Introduction.}

The commonly accepted starting point for writing down the (e,e'N) matrix
element is a Feynman diagram. The Feynman diagram consists of two vertices:
one in which a nucleon is detached from the target nucleus, and the other
which represents the electron-nucleon interaction, which scatters the
nucleon into the appropriate final state. In 1970 Gross and Lipperheide \cite
{GL} wrote down such a matrix element, and Boffi and collaborators \cite{BA}
based their work \cite{BB} on this approach. Walecka and collaborators \cite
{WA}, as well as Donnelly and co-workers\cite{BD} also start from this type
of approach. Additional refinements were gradually introduced. The optical
model, initially used to describe the nucleon-nucleus distortion \cite{BA},
was later refined by using the Random Phase Approximation for the nuclear
excitations \cite{RPA}, electron spin variables were introduced \cite{BD},
the relativistic nucleon-nucleus interaction was considered \cite{PO}, and
two-step processes (or coupled channel effects) were included \cite{BS},\cite
{K}. Many good reviews of the (e,e'N) formalism exist, the latest one being
due to Kelly \cite{KE}.

The objective of the present paper is to formulate the (e,e'\ N) reaction in
terms of a wave function expansion, rather than Feynman-diagram matrix
elements This treatment permits one a) to understand the assumptions which
underlie the coupled-channel treatment of the hadron-nucleus channels, which
are commonly inserted into the Feynman matrix elements, b) to include the
effect of the (e,e' N) channels into the treatment of dispersion corrections
to electron-nucleus scattering, c) to provide a foundation for the treatment
of the (e,e' 2N) processes.

The wave function method described here is similar in nature to the
expansions employed in three-body reaction theory \cite{GLB}, \cite{GL1}, 
\cite{GL2}. In particular, the saddle-point method is used to obtain the
asymptotic limit of the wave function for the two particles in the continuum
(the electron and the emitted nucleon), thus making contact with the matrix
element written conventionally for the Feynman diagram mentioned above.
However, the present formulation avoids the Faddeev treatment by neglecting
the electro-magnetic interaction between the two particles in the continuum,
since it is weak. In the extension to the (e,e' 2N) case, the interaction
between the two nucleons in the continuum being strong, has to be included.
This can be achieved by means by a method similar to the continuum
discretized coupled channel (CDCC) treatment\cite{NA}, \cite{JA} of the
breakup in deuteron-nucleus scattering, still without using the Faddeev
approach. However, due to the complexity of the description of the
two-nucleon correlations, both in the initial and final states, the
extension to the (e,e' 2N) case is left to a future paper.

Our treatment is as follows. The overall wave function, which contains both
the coordinates of the electron and the nucleons, is expanded on a basis set
of nuclear states which contain as much of the correlations as needed. In
the continuum, these nuclear states contain both ingoing and outgoing waves,
even though the physical boundary conditions require that only outgoing
nucleon waves exist asymptotically in the overall wave-function. This
requirement is satisfied, as is explained below. The coefficients of the
expansion of the overall wave function into this basis set are functions of
the electron coordinates, which in turn obey inhomogeneous Dirac equations.
The latter are solved by employing relativistic distorted\ Green's
functions, as is described in Appendix A. This discussion is presented in
Section 2 by means of a simple, one-nucleon example. The generalization to
the many-nucleon case is done in section 3, and the coupled channel
formulation is presented in section 4. Complications due to the
non-orthogonality between the bound states and
one-nucleon-in-the-continuum-states do not arise at this stage because in
the original expansion both the bound and the continuum nuclear states are
eigenfunctions of the same hamiltonian. This argument is not invalidated by
the subsequent replacement of many of the channels by effective complex
interactions, such as by T-matrices or optical potentials. The
non-orthogonality between the bound and the continuum optical model wave
functions is thus an artifact of the optical model approximation, but does
not invalidate the overall theory of the (e,e'N) reaction.

The treatment presented here is schematic. The electron-nucleon interaction
contains many intricate questions which have been addresses before \cite{PO}
but which are left out here. Also the details of the relativistic Dirac
description of the electron, some of which are included in Appendix A, are
left out in the text of this paper. Spin and orbital angular momentum
information is also left out. In a subsequent paper by J. Kelly on the use
of coupled equations in (e,e' p) reactions, some of the details are included
explicitly. The questions of principle addressed in this paper are: i) what
set of basis states are suitable in the description of the wave functions
involved, ii) how can one be sure that in the final state only the outgoing
parts of these wave functions survive, and iii) what are the corresponding
coefficients (the T-matrix elements) in the asymptotic waves. The
realization that only outgoing waves survive for both particles is achieved
naturally through the use of the steepest descent method in the evaluation
of a saddle point in the integration space. Such approaches are common in
the three-body literature, \cite{GL2}, \cite{NA} but the extension to the
coupled channel case requires additional thought, which is provided here.

\section{ The Basic Ideas.}

The basic ideas used in the present approach are illustrated in this section
by means of a simplified example. In this model the ''daughter'' nucleus is
described by an inert core, and the target nucleus consists of one valence
nucleon bound to this core. The interaction between the electron with the
core, and the nucleon with the core, is represented by two mean field
potentials $V_E(r_e)$ and $V_N(r_n)$, respectively. The interaction between
the electron and the hadron is denoted by $v_{en}^{(E)}$. It is commonly
expressed in the form $A_\mu j_\mu $, where $j_\mu $ is the electron current
and $A_\mu $ is the vector potential produced by the nuclear current. The
two mean field potentials are real.

The overall wave function is first expanded in a set of partial waves 
\[
\Psi (\vec r_e,\vec r_n)=\frac 1{r_er_n}\sum_{J,M_J}\psi _{J,M_J}(r_e,r_{n)}%
{\cal Y}_{J,M_J}(\hat r_e,\hat r_n), 
\]
and each radial wave is expanded into a complete set of nuclear basis states 
$\phi $ 
\begin{equation}
\psi (r_e,r_{n)}=\sum_{i=1}^B\,\chi _i(r_e)\phi _i(r_n)\,+\int_0^\infty
\,\chi _{k_Q}(r_e)\phi _Q^{(-)}(r_n)dQ.  \eqnum{2.1}
\end{equation}
In the above equation spin and angular momentum numbers $J,M_J$ should have
been included, but, they already are left out.

The discrete functions $\phi _i(r_n)$ are radial eigenstates with negative
energies of the nucleon Hamiltonian 
\begin{equation}
\left[ -\frac{\hbar ^2}{2m}\frac{d^2}{dr^2}+V_N(r)\right] \phi
_i(r)=\epsilon _i^N\phi _i(r)\,,\;\,i=1,2,...B\,,  \eqnum{2.2a}
\end{equation}
and the positive energy eigenstates form a continuum 
\begin{equation}
\left[ -\frac{\hbar ^2}{2m}\frac{d^2}{dr^2}+V_N(r)\right] \phi _Q^{(-)}(r)=%
\frac{\hbar ^2Q^2}{2m}\phi _Q^{(-)}(r)\,\;\;0\leq Q<\infty .  \eqnum{2.2b}
\end{equation}
\medskip\ The functions $\phi _Q^{(-)}(r)$ have ingoing as well as outgoing
waves. Asymptotically they are given by 
\begin{equation}
\phi _Q^{\left( -\right) }(r)\approx e^{-\,i\eta }\sin (Qr+\eta )=\frac 1{2i}%
\left( H^{\left( +\right) }(Qr)-H^{\left( -\right) }(Qr)\,e^{-2i\eta
}\right) .  \eqnum{2.3}
\end{equation}
The $H^{\left( \pm \right) }$'s are related to the outgoing and ingoing
Hankel functions according to $H^{\left( +\right) }(z)=izh^{\left( 1\right)
}(z),$ and $\,H^{\left( -\right) }(z)=-izh^{\left( 2\right) }(z)$ , where
the functions $h^{(1,2)}$ are defined in 10.1.1 of \cite{AS}, and $\eta $ is
a nuclear phase shift determined by $V_N(r)$. The orthogonality properties
of the $\phi $'s are as follows: 
\begin{equation}
<\phi _Q^{\left( -\right) }|\phi _{Q^{\prime }}^{\left( -\right)
}>=\int_0^\infty \phi _Q^{\left( +\right) }(r)\,\phi _{Q^{\prime }}^{\left(
-\right) }(r)dr=\frac \pi 2\delta (Q-Q^{\prime })  \eqnum{2.4a}
\end{equation}

\begin{equation}
<\phi _Q^{\left( +\right) }|\phi _{Q^{\prime }}^{\left( -\right)
}>=\int_0^\infty \phi _Q^{\left( -\right) }(r)\,\phi _{Q^{\prime }}^{\left(
-\right) }(r)dr=e^{-2i\eta }\frac \pi 2\delta (Q-Q^{\prime })  \eqnum{2.4b}
\end{equation}
\begin{equation}
<\phi _Q^{\left( \pm \right) }|\phi _i>=0\;;\;\;<\phi _{i^{\prime }}|\phi
_i>=\delta _{i,i^{\prime }}.  \eqnum{2.4c}
\end{equation}
In the above, 
\[
\phi _Q^{(+)}(r)=\left[ \phi _Q^{(-)}(r)\right] ^{*} 
\]
The orthogonality between the continuum states and the bound states follows
from the fact that the potential $V_N$ in Eqs. (2a) and (2b) is the same,
hence non-orthogonality is no problem here.

The functions $\chi $, coefficients of the expansion (2.1), are to be
determined from their respective Dirac Eqs., as will now be described. The
Hamiltonian of the system is 
\begin{equation}
{\cal H=\,}H_e+V_E+T_n+V_N+v_{en}  \eqnum{2.5}
\end{equation}
where $H_e$ and $T_n$ are the relativistic and non-relativistic kinetic
energy operators acting on the electron and nucleon, respectively. The
equations for the electron functions $\chi (r_e)$ are obtained by
multiplying the radial part of the overall Schr\"odinger Eq. $({\cal H}%
-E)\psi (r_e,r_n)=0$ on the left with either $<\phi _i(r_n)|$ or $<\phi
_Q^{(-)}(r_n)|$, making use of Eqs. (2.2) and (2.4), and integrating over $%
r_n$. One obtains 
\begin{equation}
(H_e+V_E-E_i^{(e)})\chi _i(r_e)=-B_i(r_e),  \eqnum{2.6a}
\end{equation}
\begin{equation}
(H_e+V_E-E_{k_Q}^{(e)})\chi _{k_Q}(r_e)=-B_Q(r_e),  \eqnum{2.6b}
\end{equation}
where the inhomogeneous driving terms $B$ are given by 
\[
B_i(r_e)=\,<\phi _i|v_{en}^{(E)}\psi (r_e,r_n)> 
\]
\begin{equation}
B_Q(r_e)=\,<\phi _Q^{(-)}|v_{en}^{(E)}\psi (r_e,r_n)>.  \eqnum{2.7}
\end{equation}
If one inserts the expansions (2.1) for $\psi (r_e,r_n)$ into Eqs. (2.7),
then Eqs. (2.6) become integro- differential equations for the functions $%
\chi _i$ and $\chi _k$. However, with the exception of the ground state term 
$\chi _g(r_e)\phi _g(r_n)$ (here $i=g)$, which is present in the incident
channel, all other $\chi $-terms are of order $v_{en}$, and hence the
inhomogeneous terms $B$ in Eqs. (2.6) are of second order in $v_{en}$, and
they can be neglected as being dispersion correction terms. We follow this
procedure here and replace $\psi (r_e,r_n)$ by $\psi _I$%
\begin{equation}
\psi (r_e,r_n)\Rightarrow \psi _I\equiv f_g^{(+)}(r_e)\phi _g(r_n), 
\eqnum{2.8}
\end{equation}
in the expressions (2.7) for the $B$'s. Here $f_g^{(+)}$ is a spinor which
satisfies 
\begin{equation}
(H_e+V_E-E_g^{(e)})f_g^{(+)}(r_e)=0.  \eqnum{2.9}
\end{equation}
It has both incident and outgoing waves and describes the scattering of an
electron from the core potential,

Regardless wether the replacement (2.8) is made or not, the solutions of
Eqs. (2.6) have, by construction, only outgoing waves asymptotically 
\begin{eqnarray}
\chi _{i} &\approx &T_{i}^{(e)}\,\exp (i\theta _{i})  \eqnum{2.10} \\
\chi _{k_{Q}} &\approx &T_{k_{Q}}^{(e)}\,\exp (i\theta _{k_{Q}})  \nonumber
\end{eqnarray}
where 
\begin{eqnarray}
\theta _{i} &=&k_{i}r_{e}+\gamma _{i}\ln (2k_{i}r_{e})  \eqnum{2.11} \\
\theta _{k_{Q}} &=&k_{Q}r_{e}+\gamma _{k_{Q}}\ln (2k_{Q}r_{e}).  \nonumber
\end{eqnarray}
In the above expression the $\gamma $'s are the Sommerfeld parameters, as
explained in Appendix A, and the angular momentum phases $l\pi /2$ have been
left out. The factors $T$ in Eqs. (2.10) are spinors which are obtained from
the solution of inhomogeneous Dirac equations, as is discussed in Appendix A.

The asymptotic expression for $\psi (r_e,r_n)$ is obtained by inserting
(2.10) into Eq. (2.1) 
\begin{eqnarray}
\psi (r_e,r_{n)} &\approx &\sum_{i=1}^B\,T_i^{(e)}\exp (\theta _i)\phi
_i(r_n)\,  \eqnum{2.12} \\
&&\ \ +\int_0^\infty \,T_{k_Q}^{(e)}\exp (\theta _{k_Q})\,\frac 1{2i}\left(
H^{\left( +\right) }(Qr_n)-H^{\left( -\right) }(Qr_n)\,e^{-2i\eta }\right)
dQ.  \nonumber
\end{eqnarray}
The electron and nucleon momenta in the expression above are related by
energy conservation, and that is what produces the correlation between the
two wave functions. Defining the total energy as $E$, the electron energy as 
$E_k^{(e)}=\sqrt{(\hbar ck)^2+(m_ec^2)^2},$ and the kinetic energy of the
nucleon as either $\epsilon _i^{(N)}$ or $E_Q^{(N)}=(\hbar cQ)^2/2m$, then
the conservation of energy equations are 
\begin{equation}
E_i^{(e)}+\epsilon _i^{(N)}=\sqrt{(\hbar ck_i)^2+(m_ec^2)^2}+\epsilon
_i^{(N)}=E  \eqnum{2.13a}
\end{equation}
\begin{equation}
E_{k_Q}^{(e)}+\epsilon _Q^{(N)}=\sqrt{(\hbar ck_Q)^2+(m_ec^2)^2}+(\hbar
cQ)^2/2m=E,  \eqnum{2.13b}
\end{equation}
from which it follows that $dQ/dk=-(k/Q)(mc^2/\epsilon _k)$. Accordingly,
the first term $\exp (i\theta _{k_Q})H^{(+)}$ in (2.12), which gives rise to 
$\exp i(r_ek+r_nQ)$, has a stationary phase point at $Q_0$, given by 
\begin{equation}
Q_0/k_0=(r_n/r_e)(mc^2/\epsilon _k^{(e)}),  \eqnum{2.14a}
\end{equation}
and the corresponding integral has a non-vanishing value which can be
evaluated by the saddle-point method \cite{AW}, while $\exp (\theta
_{k_Q})H^{(-)}$ has no such point and does not contribute. Using Eqs.
(2.13), and neglecting the rest energy of the electron in comparison to the
total energy, one obtains 
\begin{eqnarray}
Q_0 &\simeq &(r_n/r_e)mc^2/\hbar c  \eqnum{2.14b} \\
\hbar ck_0 &\simeq &E-(r_n/r_e)^2mc^2/2,  \nonumber
\end{eqnarray}
where $m$ is the nucleon mass. As described in Appendix A, the result is 
\begin{equation}
\int_0^\infty [\,T_{k_Q}^{(e)}\exp (\theta _{k_Q})\,\frac 1{2i}H^{\left(
+\right) }(Qr_n)dQ  \eqnum{2.15}
\end{equation}
\[
\approx \frac{(i)^{3/2}}2\frac{r_n^2}{r_e^{5/2}}\sqrt{2\pi mc^2/(\hbar c)}%
T_{k_0}^{(e)}\exp (\theta _{k_0})H^{(+)}(Q_0r_n)\,. 
\]

The above result shows that asymptotically the wave function contains
outgoing waves in both coordinates $r_e$ and $r_n$ in spite of the fact that
the basis set in $r_n$ contained ingoing and outgoing waves, and further,
that $r_n$ and $r_e$ are not independent of each other. Rather, the choice
of their ratio determines the value of the electron energy, and hence the
corresponding value of the nucleon energy which is to be observed. The
amplitude of the emitted electron-nucleon pair, defined in Eq. (2.10), is
determined by the T-matrix $T_k,$ which, according to Eq. (2.7) and Eq.
(A.9), is given by 
\begin{equation}
T_{k_0}^{(e)}\sim \int_0^\infty \left( f_{\kappa _0}^{(-)}(r_e)\right)
^{\dagger }\left( \phi _{Q_0}^{(-)}\,\,(r_n)\right) ^{*}v_{en}\psi
(r_er_{n)}\,dr_edr_n.  \eqnum{2.16}
\end{equation}
Here $f_{\kappa _0}$ is a Coulomb distorted electron spinor incident on the
daughter nucleus with energy corresponding to the final state, and $\psi
(r_er_{n)}$ is given by Eq. (2.8). In evaluating the integrals in Eq. (2.16)
the two variables $r_e$ and $r_n$ are independent of each other. This
expression is of the same form as given by the use of Feynman diagrams. Once
the asymptotic value of the wave function is known, the corresponding
differential cross section can be evaluated as is explained in the
literature \cite{RN}, \cite{GL3}.

As long as $\psi (r_er_n)$ contains only bound state components for the
nuclear wave function, such as in Eq. (2.8), the integral converges. If,
however, also unbound components are included, then the integral diverges, a
difficulty which has already been pointed out by Mahaux and Weidenm\"uller 
\cite{MW}. This is because now there are two particles in the continuum,
whose interaction (through $\upsilon _{en}$ in this case), extends over all
regions of space. This difficulty can be seen in a three-dimensional example
in which all continuum wave functions are replaced by plane waves. For the
case in which $\psi $ is given by Eq. (2.8), the three dimensional integral
can be done without difficulty in momentum space and yields $T_k\propto
q^{-2}\bar \phi _g({\bf q}+{\bf k}_f^n)$, where ${\bf q=k}_i^e-{\bf k}_f^e$
is the difference between the initial and final electron momenta, ${\bf k}%
_f^n$ is the momentum of the nucleon in the final state, and $\bar \phi _g$
is the Fourier transform of the bound state wave function $\phi _g/r_n$ of
the nucleon. However if $\phi _g/r_n$ is replaced by $\exp (i{\bf k}%
_n^i\cdot {\bf r}_n)$, then the corresponding result is $T_k\propto
q^{-2}\delta ({\bf q}+{\bf k}_i^n-{\bf k}_f^n)$ . This shows that the
integrals in each partial wave component are ill behaved. Thus, if
components of the wave function with two or more particles in the continuum
is to be included in the overall wave function $\psi $, (such as one
electron and one or more nucleons) and if the interaction between these
particles is to be included, then a different formulation is required. This
is commonly done in three-body calculations by going to the coordinates of
relative motion and center of mass motion of the two interacting particles,
and will be the object of a future paper on the (e,e' 2N) reaction.

The processes in which the incident electron first excites the nucleus by
promoting the valence nucleon into an excited state and then knocks it out
from there, can be included by replacing $\psi (r_er_{n)}$ in Eq. (2.16) by
the sum of terms on the right hand side of Eq. (2.1). The functions $\chi
_i(r_e)$ with $i\neq g$ would be obtained to first order in $v_{en}^{(E)}$
by the Green's function solution of Eq. (2.6a). These terms would include
the effect of dispersion corrections on the (e,e'N) reaction, but only for
the bound state excitations.

In summary, this section exhibits the main ideas of how a continuum basis
set for the emitted nucleon can be introduced, and the asymptotic electron
wave function can be obtained by a saddle point method The expression for
the (e,e'N) matrix element is identical to the usual Feynman diagram
formulation, but transcends it if virtual nuclear excitation corrections are
included

\section{Generalization to the many nucleon case, and dispersion corrections.
}

The nuclear continuum wave function basis introduced in the previous section
is unrealistic, in that the distorting potential due to the nuclear core, $%
V_N$, is real rather than complex, as would be required for an optical
potential. This difficulty will now be removed by introducing internal
degrees of freedom of the core nucleus, with the consequence that the inert
potentials $V_N$ and $V_E$ are replaced by the microscopic interactions
between the nucleons, which in turn calls for the coupled channel formalism.
The orthogonality between the positive and negative energy nuclear states
will still hold because these functions are eigenstates of the same
Hamiltonian. The subsequent introduction of effective complex interactions,
such as the optical model or effective coupling potentials, does not
invalidate the orthogonality, however, inevitable errors are introduced when
the non local effective interactions are replaced by approximate local ones.

The target nucleus has $A+1$ nucleons with coordinates ${\bf r}_n{\bf ,r}_1%
{\bf ,r}_{2,}{\bf ...r}_A$, and the coordinates of the bound nucleons, ${\bf %
r}_1{\bf ,r}_{2,}{\bf ...r}_A,$ are collectively denoted by ${\bf \xi }$.
Antisymmetrization will be ignored, the nucleon to be ejected by the
incident electron will have coordinate ${\bf r}_n$, the electron coordinate
is ${\bf r}_e$, and, as before, spin degrees of freedom, as well as angular
momentum variables are ignored.

The Hamiltonian describing the nucleons is ${\cal H}_{A+1}({\bf r}_n,{\bf %
\xi )}$, the kinetic part for the relativistic Hamiltonian of the electron
is again denoted by $H_e({\bf r}_e)$ and hence the full Hamiltonian is 
\begin{equation}
{\cal H}=H_e+v_{e,A+1}{\bf +}{\cal H}_{A+1},  \eqnum{3.1}
\end{equation}
where 
\begin{equation}
v_{e,A+1}({\bf r}_e,{\bf r}_n,{\bf \xi )=}v_{e,A}({\bf r}_e,{\bf \xi )}%
+v^{(E)}({\bf r}_e,{\bf r}_n).  \eqnum{3.2a}
\end{equation}
and 
\begin{equation}
v_{e,A}({\bf r}_e,{\bf \xi )=}\sum_{i=1}^Av^{(E)}({\bf r}_e,{\bf r}_i), 
\eqnum{3.2b}
\end{equation}
is the electromagnetic interaction of the electron with the nucleons in the
core. In the above, $v^{(E)}(e,i)$ is the microscopic electromagnetic
interaction between the electron and a nucleon which, as described in
Section 2,. contains the Coulomb as well as current terms. At this stage no
restrictive assumptions about the nature of the electromagnetic interaction
are made. If, for example the interaction with the pion cloud (the so-called
two-body currents) are included, then $v^{(E)}({\bf r}_e,{\bf r}_i)$ does
depend not only on the position and velocity of nucleon $i$ but also on the
coordinates and velocities of the other nucleons. Although this possibility
is implicit in the above expressions, it will not be written explicitly.
Likewise, the interaction of the nucleon with the core nucleons is given by 
\begin{equation}
v_{n,A}({\bf r}_n,{\bf \xi )=}\sum_{i=1}^Av^{(N)}({\bf r}_n,{\bf r}_i), 
\eqnum{3.2c}
\end{equation}
where $v^{(N)}(n,i)$ contains both nuclear and electromagnetic interactions
of nucleon $n$ with nucleon $i$.

A set of eigenstates $\Phi ^{(A+1)}$ of the $A+1$ nucleon Hamiltonian is now
introduced. They are of two types. The first type corresponds to the bound
states of the $(A+1)$ system 
\begin{equation}
{\cal H}_{A+1}\Phi _I^{(A+1)}({\bf r}_n{\bf ,\xi )=\epsilon }_I^{(A+1)}\Phi
_I^{(A+1)}({\bf r}_n{\bf ,\xi ),}  \eqnum{3.3a}
\end{equation}
while the second set represents continuum states, which obey 
\begin{equation}
{\cal H}_{A+1}\Phi _{Q,i_0}^{(A+1,-)}({\bf r}_n{\bf ,\xi )=\epsilon }%
_{Q,i_0}^{(A+1)}\Phi _{Q,i_0}^{(A+1,-)}({\bf r}_n{\bf ,\xi ).}  \eqnum{3.3b}
\end{equation}
The continuum states $\Phi _{Q,i_0}^{(A+1,-)}$ are the complex conjugate of
the states $\Phi _{Q,i_0}^{(A+1,+)}({\bf r}_n{\bf ,\xi )}$. For the latter
one nucleon is incident with momentum $Q$ on the daughter nucleus of $A$
nucleons in a state of excitation $i_0$. The outgoing channels of these
states contain one, two or more nucleons or clusters of nucleons in the
continuum. The nuclear energy ${\bf \epsilon }_{Q,i_0}^{(A+1)}$%
\begin{equation}
{\bf \epsilon }_{Q,i_0}^{(A+1)}=\frac{\hbar ^2Q^2}{2m}+{\bf \epsilon }%
_{i_0}^{(A)}+\epsilon _s  \eqnum{3.4}
\end{equation}
is composed of the kinetic energy of the free incident nucleon, the
excitation energy ${\bf \epsilon }_{i_0}^{(A)}$ of the daughter nucleus in
the state $i_0,$ and the separation energy $\epsilon _s$ of the nucleon from
the ground states of the target to that of the daughter nucleus. The recoil
energy of the daughter nucleus is not written explicitly The bound states $%
\Phi _I^{(A+1)}$ and the continuum states $\Phi _{Q,i_0}^{(A+1,-)}$ are
eigenstates of the same Hamiltonian with different energy eigenvalues, hence
they are mutually orthogonal: 
\begin{equation}
<\Phi _I^{(A+1)}\,|\Phi _{I^{\prime }}^{(A+1)}>=\delta _{I,I^{\prime }} 
\eqnum{3.5a}
\end{equation}

\begin{equation}
<\Phi _I^{(A+1)}\,|\Phi _{Q,i_0}^{(A+1,-)}>=0  \eqnum{3.5b}
\end{equation}

\begin{equation}
<\Phi _{Q,i_0}^{(A+1,-)}\,|\Phi _{Q^{\prime },i_0}^{(A+1,-)}>=\frac \pi 2%
\delta (Q-Q^{\prime }).  \eqnum{3.5c}
\end{equation}
In the above, the ket $<\Phi _{Q,i_0}^{(A+1,-)}\,|$ implies complex
conjugation.

The states $\Phi ^{(A+1)}$ are now used as a basis set into which the
complete wave function is expanded. Indicating the angular momentum algebra
only very schematically one obtains for the expansion 
\begin{eqnarray}
\Psi ({\bf r}_e,{\bf r}_n,{\bf \xi )} &=&\frac 1{r_e}\sum_{J.M_J}\left[
\sum_I\,\chi _I(r_e)\Phi _I^{(A+1)}({\bf r}_n,{\bf \xi )+}\int_0^\infty
\,\chi _{k_Q,\,i_0}(r_e)\Phi _{Q,\,i_0}^{(A+1,-)}({\bf r}_n,{\bf \xi )}%
dQ\right] _{J,M_J}  \eqnum{3.6} \\
&&\times {\cal Y}_{J,M_J}({\bf \hat r}_e,{\bf \hat r}_n,{\bf \hat \xi ),} 
\nonumber
\end{eqnarray}
where $J$ and $M_J$ are angular momentum quantum numbers and ${\cal Y}%
_{J,M_J}({\bf \hat r}_e,{\bf \hat r}_n,{\bf \hat \xi )}$ are angular and
spin functions. This expression is the many-body generalization of Eq.
(2.1).\smallskip\ 

The coupled equations for the electron wave functions will now be derived,
while the coupled equations for the hadron functions $\Phi _I^{(A+1)}$ will
be discussed in section 4. For this purpose the notation will be simplified
in what follows by replacing the subscripts $I$ by $\alpha $ , the
subscripts $(Q,i_0)$ by $\beta ,$ i.e.,$.$%
\[
\sum_I\rightarrow \sum_\alpha \;\;;\int dQ\rightarrow \sum_\beta \,, 
\]
the combined subscript $\alpha $ and $\beta $ by $\gamma $, i.e., 
\[
\sum_I+\int dQ\rightarrow \sum_\gamma \,, 
\]
and the $(-)$ superscript on the continuum states $\Phi _\beta ^{(A+1,-)}(%
{\bf r}_n,{\bf \xi )}$ is suppressed. The replacement of a continuous
variable by a discrete one is symbolic at this point, but it occurs
naturally in numerical applications when an integral over momentum is
discretized into a sum over momentum bins, as is done in the case of
break-up in deuteron-nucleus scattering\cite{NA}\cite{JA}$.$

Inserting the expansion (3.6) into the basic equation $({\cal H}-E)\Psi =0,$
dropping the angular momentum subscripts, multiplying on the left with $%
<\Phi _\gamma ^{(A+1)}|$ , integrating over all nuclear coordinates, and
making use of the orthogonality relations (3.5), one obtains the coupled
equations for the functions $\chi _\gamma $. For the discrete electron wave
functions $\chi _\alpha $ the result is 
\begin{equation}
\left( H_e-E_\alpha ^{(e)}-U_{\alpha ,\alpha }^{(e)}\right) \chi _\alpha
=-\sum_{\gamma \neq \alpha }U_{\alpha ,\gamma }^{(e)}\chi _\gamma 
\eqnum{3.7a}
\end{equation}
where 
\begin{equation}
U_{\alpha ,\gamma }^{(e)}(r_e)=\,<\Phi _\alpha ^{(A+1)}|v_{e,A+1}\Phi
_\gamma ^{(A+1)}>.  \eqnum{3.8a}
\end{equation}
The equations for the continuum wave functions $\chi _\beta $ are 
\begin{eqnarray}
&&\left( H_e-E_\beta ^{(e)}+U_{\beta ,\beta }^{(e)}\right) \chi _\beta
+\sum_\alpha U_{\beta ,\alpha }^{(e)}\chi _\alpha +\sum_{\beta ^{\prime
}\neq \beta }U_{\beta ,\beta ^{\prime }}^{(e)}\,\chi _{\beta ^{\prime }} 
\eqnum{3.7b} \\
&=&-\sum_{\beta ^{\prime }}<\Phi _\beta ^{(A+1)}|v_{e,n}^{(E)}\,\Phi _{\beta
^{\prime }}^{(A+1)}>\chi _{\beta ^{\prime }\,,}  \nonumber
\end{eqnarray}
where, 
\begin{equation}
U_{\beta ,\beta ^{\prime }}^{(e)}=\,<\Phi _\beta ^{(A+1)}|\,v_{e,A}\Phi
_{\beta ^{\prime }}^{(A+1}>  \eqnum{3.8b}
\end{equation}
\begin{equation}
U_{\beta ,\alpha }^{(e)}=\,<\Phi _\beta ^{(A+1)}|\,v_{e,A+1}\Phi _\alpha
^{(A+1)}>  \eqnum{3.8c}
\end{equation}
and the electron energies appropriate to each channel are given by 
\begin{equation}
E_\gamma ^{(e)}=E-{\bf \epsilon }_\gamma ^{(A+1)}.  \eqnum{3.9}
\end{equation}
Contrary to what was done in Eq. (3.7a), in Eq. (3.7b) the interaction $%
v_{e,n}^{(E)}$ between the two particles in the continuum was separated out
and placed on the right hand side of Eq. (3.7b). The reason is that the
matrix element of interaction between the two particles in the continuum
will diverge, unless the wave functions in the matrix element are also
distorted by that interaction. As noted before, the electron-nucleon
interactions which occur in the matrix elements (3.8) are completely
general, and include many-body currents.

Equations (3.7) are the multi-nucleon generalizations of Eqs. (2.6). The sum
over $\gamma $ in Eq. (3.7a), contains the electromagnetic coupling to the
excited bound and continuum states of the target nucleus. With exception of
the term which refers to the ground state of the target nucleus, ($\alpha
=g) $ these terms represent dispersion corrections to elastic and inelastic
electron scattering. Some of the couplings to the bound-state excitations
have been included in the literature in the calculation of dispersion
corrections to elastic electron scattering by means of explicit coupled
channel equations \cite{HC}, \cite{Ra}, but explicit coupling to the
continuum states, which contain the effect of the (e,e'N) reactions on the
elastic or inelastic electron scattering, and which arise from the terms $%
\gamma =\beta $ in the sum in Eq. (3.7a), have up to now been ignored.
Similarly, the second and third terms in the first line of Eq. (3.7b), with
the exception of the term $\alpha =g$ , also represent dispersion
corrections.

The following approximations will now be made:

a) The electromagnetic coupling between the electron and the nucleon in the
continuum will be ignored. This means that the terms in the second line of
Eq. (3.7b) will be set to zero. Should it be desired to include this
coupling, then the structure of the wave functions has to be modified, by
making a transformation to coordinates of relative motion between the two
particles in the continuum, and the coordinate of the center of mass of the
two nucleons relative to the daughter nucleus. Such transformations are
common in three-body calculations\cite{GL3}, \cite{NA}, \cite{JA}.

b) Dispersion correction terms will be ignored. This means that the terms on
the right hand side of Eq. 3.7a) are replaced by 
\begin{eqnarray*}
\sum_{\gamma \neq \alpha }U_{\alpha ,\gamma }^{(e)}\chi _\gamma &\rightarrow
&U_{\alpha ,g}^{(e)}\chi _g\;\;if\,\,\alpha \neq g \\
\sum_{\gamma \neq \alpha }U_{\alpha ,\gamma }^{(e)}\chi _\gamma &\rightarrow
&0\;\;if\,\,\alpha =g.
\end{eqnarray*}
As a result, Eq. (3.7b) together with assumption a) reduces to 
\begin{equation}
\left( H_e-E_\beta ^{(e)}+U_{\beta ,\beta }^{(e)}\right) \chi _\beta
=-U_{\beta ,g}^{(e)}\chi _g.  \eqnum{3.10}
\end{equation}

Since in Eq. (3.7b) one is dealing with continuum bins, it is not clear how
valid it is to ignore the bins which are the closest neighbors to bin $\beta
.$ This point needs further investigation. Once these terms are ignored, one
obtains Eq. (3.10), which is of the form of Eqs. (2.6 b), with $B_Q(r_e)$
given by 
\begin{equation}
B_{Q,\,i_0}=U_{\beta ,g}^{(e)}\chi _g=\,<\Phi
_{Q,\,i_0}^{(A+1,-)}|v_{e,A+1}\Phi _G^{(A+1)}>f_g^{(+)}(r_e).  \eqnum{3.11}
\end{equation}
Hence the effect of the inhomogeneous term $B_{Q,i_0}$ on $\chi _\beta $ can
be carried out in the same manner as described there. In Eq. (3.11) the
function $\chi _g$ was replaced by $f_g^{(+)},$ the solution of the Dirac
equation for a electron elastically scattered by the target nucleus in its
ground state, because, under the assumption of no dispersion corrections, $%
\chi _g$ reduces to the latter. The functions $\Phi _{Q,\,i_0}^{(A+1,-)}$ in
Eq. (3.11) is quite general, but because of the choice of the $(-)$ boundary
condition, in the limit when the electron and nucleon coordinates go to
infinity, only the nucleon channel associated with the daughter nucleus in
state $i_0$ survives in the integral term in Eq. (3.6). Nevertheless, the
presence of the other channels included in $\Phi _{Q,\,i_0}^{(A+1,-)}$
affects the value of $B_{Q,\,i_0}$, as will be discussed further in the next
section.

The assumption b) of ignoring dispersion corrections is not crucial to the
present formulation. Their contribution to the $(e,e^{\prime }N)$ matrix
element is obtained by replacing Eq. (3.11) by 
\begin{equation}
B_{Q,\,i_0}=\,\sum_\alpha <\Phi _{Q,\,i_0}^{(A+1,-)}|v_{e,A+1}\Phi _\alpha
^{(A+1)}>\chi _\alpha ^{(+)}(r_e),  \eqnum{3.12}
\end{equation}
where to first order in the electromagnetic interaction the functions $\chi
_\alpha $ are obtained from the solution of Eq. (3.7a), in which the sum
over $\gamma $ is replaced by the single term $U_{aG}^{(e)}\,f_g^{(+)}(r_e)$.

In summary, the main point of the present section is to include the
many-body degrees of freedom of the target nucleus by enlarging the meaning
of the nuclear basis functions $\Phi _I^{(A+1)}$ and $\Phi
_{Q,\,i_0}^{(A+1,-)}$ . The rigorous calculations of these functions is
impossible at this point. However, the stage is now set for making practical
approximations, by writing restricted sets of coupled equations for these
functions, as is explained in the next section.

\section{The Coupled Equations.}

The functions $\Phi _\gamma ^{(A+1)}({\bf r}_n{\bf ,\xi )}$ defined in the
previous sections are general, containing zero, one, or more nucleons (or
clusters) in the continuum. These functions are used in the expression
(3.11) for the quantities $B_{Q.i_0}$ , which in turn are required to obtain
the asymptotic value of the electron and nucleon wave functions. In the
present section an illustration is given of how to calculate the quantities $%
B_{Q.i_0}$ when the functions $\Phi _{\beta \gamma }^{(A+1,-)}$ are
restricted to a subset which contains not more than one nucleon in the
continuum. The coupled channel treatment is analogous to the customary case
of scattering of a nucleon from a target nucleus containing $A$ nucleons.
However, the present application differs from the conventional elastic
scattering treatment in that the daughter nucleus can be in a highly excited
state $i_0$ , and hence the effective interactions, such as the optical
model, will in principle be different from the case that $i_0$ represents
the ground sate. There are additional reasons why the use of an optical
model is not appropriate under these conditions, as will now be discussed.

The restricted set of functions $\Phi _\gamma ^{(A+1)}$ will be denoted by 
\[
P\Phi _\gamma ^{(A+1)} 
\]
where P is a projection operator which eliminates all components of $\Phi
_\gamma ^{(A+1)}$ which have more than one nucleon in the continuum. The
complementary projection operator is $Q$ , so that $(P+Q)\Phi _\gamma
^{(A+1)}=\Phi _\gamma ^{(A+1)}.$ The functions $P\Phi _\gamma ^{(A+1)}$ can
be expanded in terms of the bound states $\Phi _i^{(A)}$ of the daughter
nucleus 
\begin{equation}
P\Phi _\gamma ^{(A+1)}({\bf r}_n{\bf ,\xi )=}\sum_i\frac 1{r_n}\varphi
_{\gamma ,i}({\bf r}_n)\Phi _i^{(A)}({\bf \xi )}  \eqnum{4.1}
\end{equation}
where the expansion coefficients $\varphi _{\gamma ,i}({\bf r}_n)$ obey the
usual coupled equations driven by the nucleon-nucleon interaction and where
the functions $\Phi _i^{(A)}({\bf \xi )}$ are bound state eigenfunctions of\ 
${\cal H}_A$ 
\begin{equation}
{\cal H}_A\Phi _i^{(A)}({\bf \xi )=}\left( {\bf \epsilon }_i^{(A)}+\epsilon
_s\right) \Phi _i^{(A)}({\bf \xi ),\;}  \eqnum{4.2}
\end{equation}
Here ${\bf \epsilon }_i^A$ is the energy of the state above the ground
state, and $\epsilon _s$ is the separation energy of the nucleon from the
ground state of the $(A+1)$ system to that of the $A$-body system. The total
nuclear Hamiltonian is 
\begin{equation}
{\cal H}_{A+1}=T_n(r_n)+v_{n,A}(r_n,\xi )+{\cal H}_A(\xi )  \eqnum{4.3}
\end{equation}
where $T_n$ is the kinetic energy operator acting on the nucleon with
coordinate $r_n$ , and $v_{n,A}$ was defined in Eq. (3.2c).

The functions $P\Phi _\gamma ^{(A+1)}$ obey the equation 
\begin{equation}
\left( P{\cal H}_{A+1}P-E_\gamma ^{(n)}\right) P\Phi _\gamma ^{(A+1)}=0, 
\eqnum{4.4}
\end{equation}
where \cite{HF} 
\begin{equation}
P{\cal H}_{A+1}P=T_n+v_{n,A}^{(eff)}+P{\cal H}_AP,  \eqnum{4.5a}
\end{equation}
and where the effective hadron interaction is given by 
\begin{equation}
v_{n,A}^{(eff)}=Pv_{n,A}P+Pv_{n,A}Q\left( \epsilon _\gamma ^{(A+1)}-Q{\cal H}%
_{A+1}Q+i\epsilon \right) ^{-1}Qv_{n,A}P,  \eqnum{4.5b}
\end{equation}
and where 
\begin{equation}
E_\gamma ^{(n)}=\epsilon _I^{(A+1)}-\epsilon _i^{(A)}-\epsilon _s. 
\eqnum{4.5c}
\end{equation}
The interaction $v_{n,A}^{(eff)}$ differs from $v_{n,A}$ in that it includes
the effect of the channels with two or more nucleons in the continuum

The coupled equations for the expansion coefficients are obtained by
inserting the expansion (4.1) into Eq. (4.4), multiplying the equations on
the left with $<\Phi _i^{(A)}({\bf \xi )|}$, integrating over the
coordinates $\xi $ , and making use of the orthogonality properties of the
states $\Phi _i^{(A)}$. The resulting, well known, coupled equations are 
\begin{equation}
\left( -\frac{\hbar ^2}{2m}\frac{d^2}{dr_n^2}-E_{\gamma ,i}^{(n)}\right)
\varphi _{\gamma ,i}(r_n)+\sum_{i^{\prime }}\int U_{i,i^{\prime
}}^{(n)}(r_n,r_n^{\prime })\varphi _{\gamma ,i^{\prime }}(r_n^{\prime
})dr_n^{\prime }=0  \eqnum{4.6a}
\end{equation}
where 
\begin{equation}
U_{i,i^{\prime }}^{(n)}(r_n,r_n^{\prime })=<\Phi
_i^{(A)}|v_{n,A}^{(eff)}\Phi _{i^{\prime }}^{(A)}>.  \eqnum{4.6b}
\end{equation}
In the above the angular momentum terms such as $l(l+1)/r_n^2$ have again
been left out. The potential $U_{i,i^{\prime }}^{(n)}$ is non local and
complex and depends on the state $\gamma $ in view of the presence of the
Green's function in Eq. (4.5b). For negative energies $(E_{\alpha
,i}^{(n)}\leq 0)$ $\,U_{i,i^{\prime }}^{(n)}$ should become real and the
states $\varphi _{\alpha ,i}$ decay exponentially at large distances. For
positive energies $(E_{\beta ,i}^{(n)}>0)$ the states $\varphi _{\beta ,i}$
obey the (-) boundary condition, i.e., asymptotically for all states $i$
these functions are ingoing, with the exception of state $i_0$, for which
they are both ingoing and outgoing. For this case the energy $E_{\beta
,i^{\prime }}^{(n)}$ can also be expressed as 
\begin{equation}
E_{\beta ,i^{\prime }}^{(n)}=\frac{\hbar ^2Q_{i^{\prime }}^2}{2m}=\frac{%
\hbar ^2Q^2}{2m}+\epsilon _{i_0}^{(A)}-\epsilon _i^{(A)}  \eqnum{4.6c}
\end{equation}

The quantities $B_{Q,i_0},$ defined in Eq. (3.11), will now be expressed in
terms of the functions $\varphi _{\gamma ,i}$. By inserting Eqs. (4.1) into
Eq. (3.11), one obtains 
\begin{equation}
B_{Q,i_0}=\sum_i<\varphi _{\beta ,i}(r_n)\,\Phi _i^{(A)}(\xi )|\left(
v_{e,A+1}\right) \Phi _G^{(A+1)}(r_n,\xi )>f_g^{(+)}(r_e),  \eqnum{4.7}
\end{equation}
where the subscript $\beta $ stands for $(Q,i_0)$. This is the main result
of the present section. It differs from the conventional DWBA expression 
\[
B_{Q,i_0}=<f_{OM}^{(-)}(r_n)\,\Phi _{i_0}^{(A)}(\xi )|\left(
v_{e,A+1}\right) \Phi _G^{(A+1)}(r_n,\xi )>f_g^{(+)}(r_e) 
\]
in that transitions to the daughter nucleus in the final state $i_0$ can
take place by going first to a different state $i$ through the
electromagnetic coupling interaction, and from there to the final state via
the strong hadron coupling effects. The probability for the latter step is
given through the amplitude $\varphi _{\beta ,i}$ obtained from the solution
of the coupled equations (4.6a). One of these states $i$ could be an
important doorway state, whose presence would spread the (e,e'p) transition
strength to neighboring states, as long as the latter are coupled to the
doorway state \cite{O}. A formal justification for this point has apparently
not been given previously.

Examples of the use of Eq. (4.7) are as follows. a) The effect of coupling
permits the inclusion of the charge exchange process, which consists of an $%
(e,e^{\prime }n)$ transition followed by a $(n,p)$ process \cite{BS}, \cite
{K}. b) If the state $i_0$ is chosen to be highly excited, which corresponds
to a large missing energy, then the coupling to low-lying excitations such
as hole states of valence nucleons, provides a quantitative description of
how the nucleon, first emitted by knock-out from a valence orbit, excites
the daughter nucleus on its way out of the nucleus. c)\ Coupling to excited
states of the nucleon, such as a $\Delta $ excitation,.can describe the role
of nucleon excitation in the (e,e'p) process. All of these effects would be
missed in the DWBA- optical model formulation.

The number of states included in the sums over $i$ in Eqs. (4.6) and (4.7)
can be excessively large. It is usual to replace most of the states in terms
of a few ''important'' ones by means of expressions involving channel
coupling Green's functions. As a result the states which are kept explicitly
are then coupled to each other by effective interactions which are different
from those of Eq. (4.5b) because they simulate the effect of a different
channel space. They are non local, poorly known, and difficult to work with.
In addition, the electromagnetic interaction $v_{e,A+1}$ in Eq. (4.7) will
also have to be replaced by an effective interaction, 
\begin{equation}
v_{eff}^{(E)}=\left[ 1+v_{n,A}^{(eff)}{\cal G}^{\dagger }\right] v_{e,A+1} 
\eqnum{4.8}
\end{equation}
as is shown in appendix C, and as has already been discussed\cite{BO}. The
effective interaction $v_{eff}^{(E)}$ resembles a hybrid nucleon-nucleon $T-$%
matrix. It is a mixture of electromagnetic and nuclear interactions, as
shown by the occurrence in Eq. (4,8) of both the quantities $v_{n,A}^{(eff)}$
and $v_{e,A+1}$ . It is also non-local, in view of the occurrence of the
many-body channel Green's function ${\cal G}^{\dagger }$.Thus, if only one
state is used in Eq. (4.7), and that state is distorted by an optical
potential, then the electromagnetic interaction has to be changed
accordingly.

In spite of the fact that many channels other than the one being measured
can contribute to $B_{k_Q,i_0}$ in Eq. (4.7), the only channel which
survives asymptotically in the overall wave function is the channel $i_0$,
since this is the only channel which has both outgoing as well as ingoing
waves. This follows if one inserts the asymptotic form of the electron wave
function 
\[
\chi _{k_Q,i_0}\approx T_{k_Q}^{(e)}\exp (i\theta _{\kappa _Q}(r_e) 
\]
and the nucleon wave functions 
\[
\Phi _{Q,i_0}^{(A+1,-)}({\bf r}_n{\bf ,\xi )\approx }\sum_j\frac 1{r_n}\frac %
1{2i}\sum_j\left( H_{i_0}^{(+)}(r_n)\delta
_{i_0,j}-H_j^{(-)}(r_n)T_{j,i_0}^{(N)}\right) \Phi _j^{(A)}({\bf \xi )} 
\]
into the integral term in Eq. (3.6), and uses energy conservation to
correlate the electron and nucleon wave functions. The saddle-point method
of evaluating the integrals over $dQ$ then eliminates all terms but the ones
which have outgoing waves in both the electron and nucleon coordinates, and
only $H_{i_0}^{(+)}(r_n)\Phi _{i_0}^{(A)}({\bf \xi )}$ survives. One regains
the asymptotic result given by Eq. (2.15), with the only difference that the
inhomogeneous term $B_{Q,\,i_0}$, which determines the asymptotic electron
wave function factor $T_{k_Q}^{(e)}$ , is now given by the more general
expression (4.7).

\section{Summary and conclusions.}

A method has been described which both justifies and criticizes the common
practice of replacing the final state wave-function in the usual Feynman
matrix element for the (e,e' N) reaction by a coupled channel wave function.
The method points out the approximations which are implicitly made when the
final state wave-function is replaced by an optical-model distorted wave, or
when only a too restricted set of coupled equations are used. In these cases
the electromagnetic electron-nucleon interaction has to be replaced by an
effective interaction given by Eq. (4.8), which compensates for the
approximations made in the wave function. This correction goes beyond the
frequently applied Perey Factor correction, since the latter only makes up
for the omitted non-localities in the optical model potential, and not, as
Eq. (4.8) does, for the multi step transitions between channels, which
precede or follow the electromagnetic transition. Difficulties which are due
to the non-orthogonality between single particle bound states and optical
model scattering states, do not occur in the coupled channel formulation,
because the basic states of interest are eigenstates for positive and
negative energies, respectively, of the same full hadron Hamiltonian. A
further advantage of this approach is that it provides the theoretical
framework for following the flow of flux from the initial channel (the
ground state of the target nucleus) to the various segments of the inelastic
channel space, into which the (e,e'N) strength is distributed. This
information is needed to properly interpret the nuclear excitation aspect of
the transparency information, obtained from (e,e' N) or ($\mu ,\mu ^{\prime }
$ N) reactions\cite{TRANS}.

The method developed here establishes a complete wave-function framework for
describing the electron-nucleus system, so that the coupling between channels%
\cite{BS}, \cite{K} in the theory of the (e,e' N) reaction can be
justifiably included. The basis set of hadronic wave functions has ingoing
as well as outgoing nucleon waves, similar to what is the case for wave
functions used in the calculation of transition matrix elements By the use
of the saddle-point method in the integral over the continuum basis
functions in the asymptotic limit, the ingoing part of the hadronic wave
function cancels. The method is of the type commonly used in three-body
calculations, but does not use the rigorous Faddeev formalism since the
electromagnetic interaction between the two particles in the continuum is
neglected.

The formalism is applicable, with several important modifications, to the
description of the emission of two nucleons from the target nucleus. The
interactions between the two nucleons in the continuum is generally ignored 
\cite{RY}, and in order to include them a procedure can be introduced which
has strong similarities to the description of the effects of deuteron
breakup on deuteron-induced nuclear reactions \cite{NA}, \cite{JA}.
Experiments already point \cite{B} to the influence of the (e,e' 2N) process
in the emission of one nucleon, and hence a theory for this process is
timely.

In summary, a derivation was given of the coupled channel wave-function
description of the emission of a nucleon from a target nucleus by an
incident electron. The derivation enables one to make contact with the
theories of the optical model or of the coupled channel description of
nucleon-nucleus reactions, and is generalizable to the case of the (e,e'2N)
process.

\bigskip\ \bigskip\ 

{\bf Acknowledgments}.

The author is grateful for conversations with Dr. J. J. Kelly, at the
University of Maryland, Dr. S. Krewald at the Kernforschungs Anlage in
J\"ulich, and several years ago, with Dr. T. W. Donnelly at MIT, concerning
the feasibility of incorporating the coupled channel formalism into the
description of (e,e' N) reactions.\newpage\ 

{\bf Appendix A:\newline
Asymptotic solution of the inhomogeneous Dirac Eq.}

If the Dirac eq. describing the scattering on an electron from the
electrostatic field of a nucleus, has also an inhomogeneous driving term,
then the solution can be written in terms of a relativistic distorted
Green's function acting on the inhomogeneous terms. The purpose of this
Appendix is to present expressions for the asymptotic form of the result.,
in two different forms. 1. Involving the relativistic spinors which are the
regular and irregular solutions of the distorted homogeneous Dirac equation,
and 2. Involving the solutions of the Schr\~odinger-like solutions of the
second order equations which are equivalent to the first order Dirac
equation. This second result is of value when a comparison between the
relativistic and non-relativistic treatments of the distortion of the
nucleons in the (e,e'N) reaction is desired.

Following the notation of Mott and Massey,\cite{MM} p. 228, a Dirac spinor
of a scattering wave function, when expanded into partial waves, has two
sets of radial functions for the upper and lower components. They are
denoted by $G_l(r)$ and $F_l(r)$ for one set of solutions, and $G_{-l-1}(r)$
and $F_{-l-1}(r)$ for the other set. Each of these still multiplies a
two-component spinor of the spherical harmonic angular functions, as
indicated in Ref.\cite{MM}. If one defines new radial functions ${\cal G}%
_\lambda (r)=G_\lambda (r)\times (kr)$ and ${\cal F}_\lambda (r)=F_\lambda
(r)\times (kr),$ with $\lambda =l$ or $-l-1$, then the coupled radial Dirac
equations are 
\begin{equation}
a_{-}{\cal G}_\kappa (r)-(\frac d{dr}-\frac \kappa r){\cal F}_\kappa (r)=0 
\eqnum{A.1a}
\end{equation}
\begin{equation}
a_{+}{\cal F}_\kappa (r)+(\frac d{dr}+\frac \kappa r){\cal G}_\kappa (r)=0 
\eqnum{A.1b}
\end{equation}
where 
\begin{equation}
a_{\pm }=[E\pm mc^2\pm V-e\Phi ]/\hbar c.  \eqnum{A.2}
\end{equation}
In the above, the quantum numbers $\kappa $ are the eigenvalues of the
operator $K=-(\sigma \cdot L+\hbar )$, which assume positive or negative
integer values. They are related to the angular momentum quantum numbers $l$
and the index $\lambda $ as follows. For $\kappa >0,$ $\lambda =-l-1$ and $%
\kappa =l$. For $\kappa <0,$ $\lambda =l$ and $\kappa =-l-1$. The vector
spherical harmonics ${\cal Y}_{\kappa ,\mu }$ which multiply the functions $%
{\cal G}$ and ${\cal F}$ are eigenfunctions of the operator $K.$ The two
upper and lower components of the Dirac spinors are given by $r^{-1}{\cal G}%
_k(r){\cal Y}_{\kappa .\mu }$ and -$r^{-1}i{\cal F}_{-\kappa ,\mu }{\cal Y}%
_{-\kappa ,\mu }$, respectively \cite{TH}. In the definition of $a_{\pm }$, $%
E$ is the total energy of the electron, $mc^2$ is its rest energy, $V$ is
the is the scalar part of the potential, which is added to the rest mass and
which is present in the case of a nucleon scattering from the field of a
nucleus, and $e\Phi $ is the electrostatic Coulomb potential which is the
vector part of the potential, added to the energy. The latter is negative
(positive) if electrons (protons) interact with the positive charge
distribution of the target nucleus. The asymptotic values of ${\cal G}$ are 
\begin{equation}
{\cal G}_\kappa \approx \sin (\phi _l+\eta _\kappa )\,\exp (i\eta _\kappa ) 
\eqnum{A.3a}
\end{equation}
where 
\begin{equation}
\phi _l=kr-l\pi /2+\gamma \ln (2kr).  \eqnum{A.3b}
\end{equation}
The value of $l$ in Eq.(A.3b) is related to $\kappa $ as described above, $%
\gamma $ is the Sommerfeld parameter $Ze^2/\hbar v$, and $k$ is the wave
number, related to the energy according to $\sqrt{(\hbar ck)^2+(mc^2)^2}=E$.

The inhomogeneous equations to be solved are 
\begin{equation}
a_{-}{\cal P}_\kappa (r)-(\frac d{dr}-\frac \kappa r){\cal Q}_\kappa
(r)=b_\kappa ^{(U)}(r)  \eqnum{A.4a}
\end{equation}
\begin{equation}
a_{+}{\cal Q}_\kappa (r)+(\frac d{dr}+\frac \kappa r){\cal P}_\kappa
(r)=b_\kappa ^{(L)}(r)  \eqnum{A.4b}
\end{equation}
Here ${\cal P}$ and ${\cal Q}$ replace the functions ${\cal G}$ and ${\cal F}
$ in the homogeneous Dirac equation and $b^{(U)}$ and $b^{(L)}$ are the
upper and lower components, respectively, of the inhomogeneous driving term.
Asymptotically thes functions ${\cal P}$ and ${\cal Q}$ have only outgoing
terms, the coefficients of which define the T-matrix elements $T_p$ and $T_q$%
\begin{equation}
{\cal P}_\kappa \simeq T_{{\cal P}\kappa }\exp (i\phi _l)  \eqnum{A5a}
\end{equation}
\begin{equation}
{\cal Q}_\kappa \simeq T_{{\cal P}\kappa }\exp (i\phi _l).  \eqnum{A5a}
\end{equation}

Following the expressions given by Onley\cite{DO}, for the relativistic
Green's function, one obtains for the solution of Eq.\ (A4) the result 
\begin{equation}
{\cal P}(r)={\cal G}_R(r)A(r)+{\cal G}_I(r)B(r)  \eqnum{A.6a}
\end{equation}
and 
\begin{equation}
{\cal Q}(r)={\cal F}_R(r)A(r)+{\cal F}_I(r)B(r)  \eqnum{A.6b}
\end{equation}
where 
\begin{equation}
A(r)=\int_r^\infty \left( {\cal G}_Ib^{(U)}+{\cal F}_Ib^{(L)}\right)
dr^{\prime }  \eqnum{A.7a}
\end{equation}
and 
\begin{equation}
B(r)=\int_0^r\left( {\cal G}_Rb^{(U)}+{\cal F}_Rb^{(L)}\right) dr^{\prime }.
\eqnum{A.7b}
\end{equation}
In the above the pair of functions $({\cal G}_R,{\cal F}_R)$ are the regular
solutions of Eqs. (A.1). with asymptotic behavior for ${\cal G}_R$ given by
(A.3a). Similarly, the functions $({\cal G}_I,{\cal F}_I)$ are the irregular
solutions, with ${\cal G}_I$ normalized such that the asymptotic value is
given by 
\begin{equation}
{\cal G}_I\approx -\frac{E+mc^2}{\hbar ck}\exp (i\phi _l).  \eqnum{A.8}
\end{equation}
The asymptotic behavior of the corresponding functions ${\cal F}$ is
obtained in terms of the ${\cal G}$ 's from Eq. (A.1b). The relativistic
equivalent of the Wronskian between the regular and irregular solutions is 
\[
{\cal F}_R{\cal G}_I{\cal -F}_I{\cal G}_R=1 
\]
whose value is a constant, which is equal to unity for the choice of the
normalizations of ${\cal G}_R$and ${\cal G}_I$ as described above.

As a result of the above, the asymptotic value of the spinor $({\cal P}$,$%
{\cal Q}{\frak )}$, defined in Eq. (A.6), is of the form given by Eq. (A.5),
with the spinor 
\begin{equation}
\left( 
\begin{array}{c}
T_{{\cal P\kappa }} \\ 
T_{{\cal Q\kappa }}
\end{array}
\right) =T_\kappa \left( 
\begin{array}{c}
-(E+mc^2)/(\hbar ck) \\ 
i
\end{array}
\right)  \eqnum{A.9a}
\end{equation}
and 
\begin{equation}
T_\kappa =B(\infty )=\int_0^\infty \left( {\cal G}_Rb^{(U)}+{\cal F}%
_Rb^{(L)}\right) dr.  \eqnum{A.9b}
\end{equation}
This is the main result of the Appendix.\bigskip\ 

The asymptotic expressions for the solution of the inhomogeneous Dirac
equation will now be given in terms of the solution of the equivalent second
order Schr\~odinger-like equation. This procedure consists in transforming
the two coupled first order differential Dirac equations for the radial
partial wave functions\ (with or without the driving terms) into uncoupled
second order differential equations, to which the usual wronskian techniques
are then applied in order to extract the coefficient of the asymptotic
outgoing wave. The ''Darwin factors'' (akin to Perey factors) which are
required in this transformation (in order to eliminate the first order
differential terms) in the end cancel away in the term which resembles the
non-relativistic expression, but remain for the additional terms which are
also present.

By differentiation and linear transformations, the first order coupled
differential equations (A.1) and (A.5) are reduced to uncoupled differential
equations of first and second order for ${\cal F}$ and ${\cal G},$ and
similarly for ${\cal P}$ and ${\cal Q}.$ The coefficients in front of the $%
{\cal G}^{\prime \prime },{\cal G}^{\prime }$ and ${\cal G}$ are the same as
the coefficients in front of ${\cal P}^{\prime \prime },{\cal P}^{\prime }$
and ${\cal P}$ but in the equations for the latter the inhomogeneous terms
and their derivatives also appear. Similarly for ${\cal F}$ and ${\cal Q}$.
The first order terms of these equations are then eliminated by introducing
the Darwin factors 
\begin{equation}
F_{\pm }(r)=\sqrt{a_{\pm }(r)}  \eqnum{A.10}
\end{equation}
and by performing the transformation 
\begin{equation}
{\cal G}(r)=F_{+}(r)\,g(r)\;;\;{\cal F}(r)=F_{-}(r)\,f(r)  \eqnum{A.11a}
\end{equation}
\begin{equation}
{\cal P}(r)=F_{+}(r)\,p(r)\;;\;{\cal Q}(r)=F_{-}(r)\,q(r).  \eqnum{A.11b}
\end{equation}
one obtains second order differential equations for $g$ and $f$, and
similarly $p$ and $q$, with unit coefficients in front of the second-order
derivative terms. These equations for $p$ and $q$ of course contain the
inhomogeneous terms, which are denoted $b_p$ and $b_q$, respectively, but
otherwise they are identical to the equations for $g$ and $f$. These
transformations were already given by Darwin \cite{MM}, and are routinely
used in the literature\cite{BC} in order to cast the Dirac equations into
Schr\"odinger form.

One obtains the asymptotic factor $T_{{\cal P}}$, defined in Eq.\ (A.5), by
multiplying the equation for $p^{\prime \prime }$ with $g$ and the equation
for $g^{\prime \prime }$ with $p$, and integrating the difference over $r$
from 0 to $\infty .$ Because the coefficients in front of the $g$ and $p$
are the same, these terms cancel, and by making use of the asymptotic
relations (A.3) and (A.5) one obtains 
\begin{equation}
\int_0^\infty \left( gp^{\prime \prime }-pg^{\prime \prime }\right)
dr=-kT_p\left( F_{+}(\infty )\right) ^{-2}.  \eqnum{A.12a}
\end{equation}
The left hand side of the above equation is also equal to $\int_0^\infty gb_p%
{\cal \,}dr.$ Combining these two results, one finally obtains, 
\begin{eqnarray}
T_p &=&\frac{E+mc^2}{pc}\int_0^\infty {\cal G\,}b^{(U)}\,dr  \eqnum{A.13a} \\
&&+\frac{E+mc^2}{pc}\int_0^\infty {\cal G}\left[ \left( \frac{a_{+}^{\prime }%
}{a_{+}}+\frac \kappa r\right) b^{(L)}/a_{+}-b^{(L)}\,^{\prime
}/a_{+}\right] dr  \nonumber
\end{eqnarray}

By proceeding in a similar way for ${\cal F}$ and ${\cal Q}$, one obtains 
\begin{equation}
\int_0^\infty \left( fq^{\prime \prime }-qf^{\prime \prime }\right) dr=-i\,k%
\frac{\hbar ck}{E+mc^2}T_q\left( F_{-}(\infty )\right) ^{-2}=-iT_q 
\eqnum{A.12b}
\end{equation}
The left hand side of the above equation is also equal to $\int_0^\infty fb_q%
{\cal \,}dr.$ Combining these two results, one finally obtains, 
\begin{eqnarray}
T_q &=&i\int_0^\infty {\cal F\,}b^{(L)}\,dr  \eqnum{A.13b} \\
&&-i\int_0^\infty {\cal F}\left[ \left( \frac{a_{-}^{\prime }}{a_{-}}-\frac %
\kappa r\right) b^{(U)}/a_{-}-b^{(U)}\,^{\prime }/a_{-}\right] dr.  \nonumber
\end{eqnarray}

The first lines are the same as what one gets in the non-relativistic
treatment, however, the functions which enter into these expressions are the
solutions ${\cal G}$ or ${\cal F}$ of the Dirac equation. If one used the
non-relativistic simulations $g$ or $f$of these functions, one would first
have to multiply them by the normalized Darwin factors 
\[
\bar F_{\pm }(r)=\sqrt{a_{\pm }(r)/a_{\pm }(\infty )} 
\]
in order to transform them into the functions ${\cal G}$ or ${\cal F}$.
These factors are many times confused with the Perey damping factors,
because they have approximately the same radial dependence\cite{GR}. However
these two type of factors have a different physical origin and should not be
confused with each other. The terms in the second lines in Eqs. (A.13) are
related to the spin-orbit potentials, since the term $\kappa /r$ is angular
momentum dependent, and the factors $a_{\pm }^{\prime }/a_{\pm }$ are the
coefficients of the spin-orbit potentials in the Schr\"odinger form of the
radial Dirac equations.

{\bf Appendix B:\newline
Asymptotic behavior via the method of steepest descent.}\smallskip\ 

In section 2 the wave function $\psi (r_e,r_n)$ was represented in terms of
an integral over a continuous set of basis functions $\phi _Q^{(-)}(r_n)$,
and the coefficients of the expansion, $\chi _{k_Q}(r_e)$, were obtained
from the solution of an in$\hom $ogeneous Dirac equation with outgoing wave
boundary conditions, $\exp (ik_Qr_e)$. Asymptotically the function $\phi
_Q^{(-)}(r_n)$ has both ingoing and outgoing waves $\exp (\pm iQ\,r_n)$, and
hence the asymptotic form of $\psi (r_e,r_n)$ has terms of the form $\exp
(ik_Qr_e\pm iQ\,r_n)$. The two continuum variables $k_Q$ and $Q$ are related
by energy conservation, Eq. (2.5b). As $Q$ increases, $k_Q$ decreases, and
hence $ik_Qr_e+iQ\,r_n$ has a saddle point at $Q_0$ where the integral over $%
Q$ has a stationary phase The integral has a finite value which can be
calculated by means of a steepest descent method around $Q_0$. On the other
hand, the integral with $\exp (ik_Qr_e-iQr_n)$ has no stationary phase and
it vanishes. The integral involving $\exp (ik_Qr_e+iQr_n)$ can be obtained
from the steepest descent method which gives [Arfken and Weber] the general
result 
\begin{eqnarray}
I(s) &\equiv &\int_Cg(z)\exp (s\,f(z)\,dz  \eqnum{B.1} \\
&\approx &\left( \frac{2\pi }{|sf^{\prime \prime }(z_0)}\right)
^{1/2}g(z_0)\exp \left( sf(z_0)+i\alpha \right) .  \nonumber
\end{eqnarray}
Here $z=Q,$ 
\begin{equation}
s\,f(z)=ik_Qr_e+iQ\,r_n,  \eqnum{B.2}
\end{equation}
$g(z)$ is a slowly varying function of $z,$ $\alpha $ is a phase which
depends on the direction of the path across the saddle point in the complex
plane, and $s$ is the variable which becomes asymptotically large. The
saddle point is determined from the requirement $df/dz=0,$ which in our case
gives 
\begin{equation}
\frac{Q_0}{k_0}=\frac{mc^2}{\epsilon _{k_0}}x  \eqnum{B.3}
\end{equation}
where $\epsilon _k$ is the energy of the electron, as defined near Eq.
(2.5), $m$ is the mass of the nucleon, and $x$ is the ratio 
\begin{equation}
x=\frac{r_n}{r_e}.  \eqnum{B.4}
\end{equation}
Conservation of energy, Eq. (2.5b), was used in the derivation of Eq. (B.3).
From the latter one can obtain $Q_0$ and $k_0$ separately. Neglecting the
rest-mass energy of the electron relative to $\epsilon _k$ one finds 
\begin{equation}
\hbar cQ_0\simeq mc^2x  \eqnum{B.5a}
\end{equation}
and 
\begin{equation}
\hbar ck_0\simeq E-mc^2x^2/2,  \eqnum{B.5b}
\end{equation}
where E is the combined energy of the electron and nucleon, as given in Eq.
(2.5b). This result serves to fix the ratio of distances $x$. If the
electron energy $\hbar ck_0$ is determined from experiment, then the
corresponding value of $x$ is determined from Eq. (B.5b), and the
corresponding value of $Q$ from Eq. (B5.a). Inserting the value of 
\begin{equation}
\frac{d^2sf}{dQ^2}=-i\frac{\hbar c}{mc^2}r_e  \eqnum{B.6}
\end{equation}
into Eq. (B.1), one obtains 
\begin{equation}
I(s)\approx g(Q_0)\left( \frac{2\pi imc^2}{\hbar c\,r_e}\right) ^{1/2}\exp
\left[ ir_e(k_0+\frac{mc^2}{\hbar c}x^2)\right] .  \eqnum{B.7}
\end{equation}
In the above, $r_e$ is assumed to be the independent variable, and $r_n$
related to it via $x$. The exponent of Eq. (B.7) can also be written as $%
\exp \left[ ir_ek_0+ir_nQ_0)\right] .$\smallskip\ 

\appendix{\bf Appendix C. Reduction of the number of channels in terms of
effective interactions.}

The sums over $i$ in Eqs. (4.6) and (4.7) contain a very large number of
terms, which would be impossible to include in numerical calculations.
Instead, it is more practical to eliminate most of the channels $i$ in terms
of a few leading channels by means of channel Greens functions and effective
interactions\cite{HF}. . In this Appendix the resulting modifications to the 
$(e,e^{\prime }p)$ formulation will be briefly indicated.

In order to simplify the notation, the potential $v_{n,A}^{(N)}$ will be
replace by $v,$ the matrix elements $<\Phi _i^{(A)}|v_{n,A}^{(N)}\Phi
_{i^{\prime }}^{(A)}>$ by $<i|v|i^{\prime }>$, and the wave functions $\phi
_{Q,i}^{(i_0,-)}(r_n)$ will be replaced by $\phi _i(r_n)$. Two channels will
be kept explicitly, and they will be denoted by $a$ and $b$. It is assumed
that the incident wave is in channel $a$. The kinetic energy operator $%
-(\hbar ^2/2m)\,(d^2/dr_n^2)$ is denoted by $T$, and the channel energy $%
E_{Q,i}$ is denoted by $E_i$. The coupled equations (4.4b) then take the
form 
\begin{equation}
(T-E_i)\phi _i+\sum_{i^{\prime }}<i|v|i^{\prime }>\phi _{i^{\prime
}}=-<i|v|a>\phi _a-<i|v|b>\phi _b  \eqnum{C.1a}
\end{equation}
\begin{equation}
(T+<a|v|a>-E_a)\phi _a+\sum_{i^{\prime }}<a|v|i^{\prime }>\phi _{i^{\prime
}}=-<a|v|b>\phi _b  \eqnum{C1.b}
\end{equation}

\begin{equation}
T+<b|v|b>-E_b)\phi _a+\sum_{i^{\prime }}<b|v|i^{\prime }>\phi _{i^{\prime
}}=-<b|v|a>\phi _a  \eqnum{C.1c}
\end{equation}
Channels $i$ can be eliminated formally \cite{HF} by using the
channel-coupled Greens function $G_{i,i^{\prime }}(r_n,r_n^{\prime }),$
defined for example in Ref. \cite{GR2}. By using these functions in Eq.
(C.1a) one obtains 
\begin{equation}
\phi _i=\sum_{i^{\prime }}G_{i,i^{\prime }}\left[ <i^{\prime }|v|a>\phi
_a+<i^{\prime }|v|b>\phi _b\right] ,  \eqnum{C.2}
\end{equation}
where integration over the second Green's function variable is implied. Upon
inserting this result into Eqs. (C.1a) and (C.1b), one obtains the
''reduced'' set of coupled equations 
\begin{equation}
(T+U_{aa}-E_a)\phi _a=-U_{ab}\phi _b  \eqnum{C.3a}
\end{equation}
\begin{equation}
(T+U_{bb}-E_b)\phi _b=-U_{ba}\phi _a,  \eqnum{C.3b}
\end{equation}
where the effective potentials $U_{\alpha \beta }$ are given by 
\begin{equation}
U_{\alpha \beta }(r_n,r_n^{\prime })=<\alpha |v|\beta >_{r_n}\delta
(r_n-r_n^{\prime })+\sum_{i,i^{\prime }}\,<\alpha |v|i>_{r_n}G_{i,i^{\prime
}}(r_n,r_n^{\prime })<i^{\prime }|v|\beta >_{r_n^{\prime }.}  \eqnum{C.4}
\end{equation}
Here $\alpha $ or $\beta $ take on the values $a$ and $b$. Because of the
presence of the $G_{i,i^{\prime }}$the potentials $U$ are complex and
non-local, and integration over the non-localities is implied in Eqs. (C.3).
One can summarize the result of Eq. (C.4) by defining an effective
nucleon-nucleon potential 
\begin{equation}
v_{eff}=v+v{\cal G}(r_n,\xi ;r_n^{\prime },\xi ^{\prime })v_{.}  \eqnum{C.5a}
\end{equation}
where the many body channel Green's function is given by 
\begin{equation}
{\cal G}(r_n,\xi ;r_n^{\prime },\xi ^{\prime })=\sum_{i,i^{\prime
}}|\,i>_\xi G_{i,i^{\prime }}(r_n,r_n^{\prime })<i^{\prime }|_{\xi ^{\prime
}}.  \eqnum{C.5b}
\end{equation}
If one makes use of the result 
\begin{eqnarray}
\phi _a^{*} &<&a|+\phi _b^{*}<b|+\sum_i\phi _i^{*}<i|=  \eqnum{C.6} \\
&&\phi _a^{*}\left[ <a|+<a|v{\cal G}^{\dagger }\right] +\phi _b^{*}\left[
<b|+<b|v{\cal G}^{\dagger }\right]  \nonumber
\end{eqnarray}
one can rewrite the expression (4.6) for $B$ in terms of that set of channel
functions $a$ and $b$ which are kept explicitly as 
\begin{equation}
B_{k_Q,i_0}=<\left[ \phi _{Q,a}^{(i_0,-)}(r_n)\,\Phi _a^{(A)}(\xi )+<\phi
_{Q,b}^{(i_0,-)}(r_n)\,\Phi _b^{(A)}(\xi )\right] |v_{eff}^{(E)}\Phi
_G^{(A+1)}(r_n,\xi )>f_G(r_e).  \eqnum{C.7}
\end{equation}
where 
\begin{equation}
v_{eff}^{(E)}=\left[ 1+v_{n,A}^{(N)}{\cal G}^{\dagger }\right]
v_{e,A+1}^{(E)}  \eqnum{C.8}
\end{equation}
This result shows that if only a few states are used in Eq. (4.6) for the
calculation of the inhomogeneous term $B$, such as for example the single
measured state $|i_0>$ times the corresponding optical model wave function,
then, in order to make up for the contribution from the other states an
effective electromagnetic interaction has to be used . This interaction is a
mixture of electromagnetic and nuclear interactions, as shown by the
occurrence of $v_{n,A}^{(N)}{\cal G}^{\dagger }$ in Eq. (C.8).\newpage\ 

{\bf REFERENCES.}

\end{document}